\documentclass[aps,prl,twocolumn,superscriptaddress]{revtex4-2}

\usepackage{graphicx}
\usepackage{lineno}
\usepackage{amsmath}
\usepackage{amssymb}
\usepackage{xcolor}

\newcommand*\regionAmsMathEnvironmentForLineno[1]{%
	\expandafter\let\csname old#1\expandafter\endcsname\csname #1\endcsname
	\expandafter\let\csname oldend#1\expandafter\endcsname\csname end#1\endcsname
	\renewenvironment{#1}%
	{\linenomath\csname old#1\endcsname}%
	{\csname oldend#1\endcsname\endlinenomath}}%
\newcommand*\regionBothAmsMathEnvironmentsForLineno[1]{%
	\regionAmsMathEnvironmentForLineno{#1}%
	\regionAmsMathEnvironmentForLineno{#1*}
}%
\AtBeginDocument{%
	\regionBothAmsMathEnvironmentsForLineno{equation}%
	\regionBothAmsMathEnvironmentsForLineno{align}%
	\regionBothAmsMathEnvironmentsForLineno{flalign}%
	\regionBothAmsMathEnvironmentsForLineno{alignat}%
	\regionBothAmsMathEnvironmentsForLineno{gather}%
	\regionBothAmsMathEnvironmentsForLineno{multline}%
}

\begin{document}


\title{Sharp transition to strongly anomalous transport in unsaturated porous media} 


\author{Andr\'{e}s Vel\'{a}squez-Parra}
\affiliation{Department of Water Resources and Drinking Water, Swiss Federal Institute of Aquatic Science and Technology, Eawag, D\"{u}bendorf, Switzerland}
\affiliation{Department of Civil, Environmental and Geomatic Engineering, ETH Z\"{u}rich, Z\"{u}rich, Switzerland}

\author{Tom\'{a}s Aquino}
\affiliation{Univ. Rennes, CNRS, G\'eosciences Rennes, UMR 6118, 35000 Rennes, France}

\author{Matthias Willmann}
\affiliation{Department of Civil, Environmental and Geomatic Engineering, ETH Z\"{u}rich, Z\"{u}rich, Switzerland}

\author{Yves M\'{e}heust}
\affiliation{Univ. Rennes, CNRS, G\'eosciences Rennes, UMR 6118, 35000 Rennes, France}

\author{Tanguy Le Borgne}
\affiliation{Univ. Rennes, CNRS, G\'eosciences Rennes, UMR 6118, 35000 Rennes, France}

\author{Joaqu\'{i}n Jim\'{e}nez-Mart\'{i}nez}
\email[]{joaquin.jimenez@eawag.ch / jjimenez@ethz.ch}
\affiliation{Department of Water Resources and Drinking Water, Swiss Federal Institute of Aquatic Science and Technology, Eawag, D\"{u}bendorf, Switzerland}
\affiliation{Department of Civil, Environmental and Geomatic Engineering, ETH Z\"{u}rich, Z\"{u}rich, Switzerland}


\date{\today}

\begin{abstract}


The simultaneous presence of liquid and gas in porous media increases flow heterogeneity compared to saturated flows. However, so far the impact of saturation on flow statistics and transport dynamics remained unclear. Here, we develop a theoretical framework that captures the impact of flow reorganization on the statistics of pore-scale fluid velocities, due to the presence of gas in the pore space, which leads to the development of a highly-structured flow field. Preferential flow is distributed spatially through the denoted {\em backbone} and flow recirculation occurs in dead-end regions branching from it. This induces a marked change in the scaling of the velocity PDF compared to the saturated case, and a sharp transition to strongly anomalous transport. We develop a transport model based on the continuous time random walk theory that successfully predicts advective transport dynamics for all saturation degrees. Our results provide a new modelling framework linking phase heterogeneity to flow heterogeneity and to transport in unsaturated media.
\end{abstract}


\maketitle


Unsaturated porous media, where liquid and gas phases coexist, play a central role in a broad range of environmental~ \cite{Lahav2010,Sebilo2013,Bouwer2002} and industrial~\cite{Panfilov2010,Winograd1981,Barbier2002} applications. Under saturated conditions, i.e., for single-phase flow, structural heterogeneity in the solid phase leads to anomalous transport~\cite{DeAnna2013,LeBorgne2011,Holzner2015,Morales2017, Moroni2007,Kang2014,Stoop2019}. This typically translates to early solute arrival and longer tailing at a given control plane, as well as non-Fickian scaling of spatial solute spreading~\cite{Bijeljic2011}. In unsaturated porous media, the presence of several immiscible or partially-miscible fluid phases in the pore space induces complex flow topologies, increasing flow tortuosity and resulting in more extreme high and low velocities~\cite{DeGennes1983, Jimenez-Martinez2017, LeBorgne2008}. However, the consequences of this heterogeneity to solute transport properties are controversial. Both an increase~\cite{Bromly2004,Haga1999,Padilla1999} and a decrease~\cite{Birkholzer1997,Vanderborght2007} of dispersion with decreasing saturation have been reported.

Here, we use images from millifluidic experiments and pore-scale numerical simulations to derive a new theoretical framework linking parameters characteristic of the medium structure and saturation degree (fraction of the pore volume occupied by the liquid) to the probability density function (PDF) of both flow rates through pore throats and velocities, and to anomalous transport dynamics. We propose a model built on the partition of the pore space into two contrasting structures, a backbone of preferential flow paths, and dead-end regions of low velocity. The model captures the abrupt change the presence of dead-ends exerts on the velocity PDF scaling for unsaturated systems compared to fully saturated conditions. Based on this theory, we predict the emergent anomalous advective transport in unsaturated systems based on a continuous time random walk (CTRW) approach.
 
We employ experimental images of a horizontal quasi two-dimensional (2D) porous medium characterizing the arrangement of two immiscible phases (water and air) under different saturation degrees~\cite{Jimenez-Martinez2017} and simulate flow at the pore scale. Four saturation degrees, $S_{\mathrm{w}}=1.00$, $0.83$, $0.77$, and $0.71$, are analyzed. Experiments were performed for low capillary numbers, ranging from $0.763$ for $S_{\mathrm{w}}=0.83$ to $0.892$ for $S_{\mathrm{w}}=0.71$. Therefore, viscous forces posed by liquid flow did not exceed the capillary resistance of the air clusters (non-wetting phase), remaining immobile~\cite{Tang2019}. The length of the system is $132\,\mathrm{mm}$, its width $87\,\mathrm{mm}$, and its thickness (vertical gap) $h=0.5\,\mathrm{mm}$. The average pore throat width (shortest distance between grains) $a_{\mathrm{m}}=1.17\,\mathrm{mm}$, and the mean pore size (meeting point of pore throats) $\lambda=1.85\,\mathrm{mm}$, leading to a porosity of $0.71$, which is typical of 2D systems~\cite{Andrade1997,Karadimitriou2016,Tallakstad2009}. 

We numerically simulate 2D steady-state Stokes (i.e., low Reynolds number) flow, in which the flow of water around the solid grains and air bubbles, extracted from experimental images, is exclusively controlled by viscous dissipation. The effect of the third dimension on depth-averaged flow is introduced in the Stokes equation through a Darcy-like term~\cite{Ferrari2015} representing the drag force exerted on the liquid by the upper and lower walls in the experimental configuration~\cite{Jimenez-Martinez2017}. A constant flow rate of $1.375\,\mathrm{mm^{3}/s}$ for the saturated case and $0.277\,\mathrm{mm^{3}/s}$ for the unsaturated cases is imposed at the inlet~\cite{Jimenez-Martinez2017}. The pixel size is $0.04$ mm $\simeq \lambda / 50$. Atmospheric pressure is imposed at the outlet. We assign a slip boundary condition (i.e., no tangential stress) to liquid--gas boundaries, and a no-slip boundary condition (i.e., zero velocity) to solid--liquid interfaces. 
\begin{figure*}
\centering
	\includegraphics{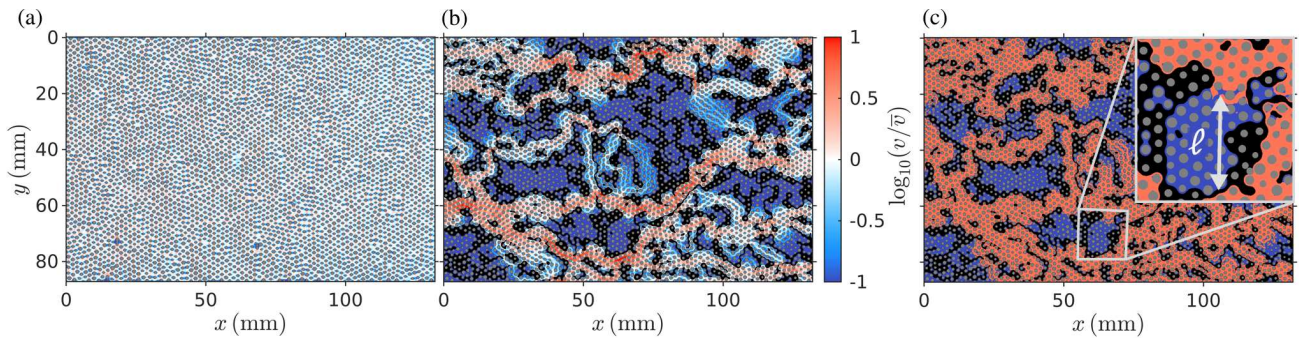}%
	\caption{\label{fig:velfield} (color online). Velocity fields obtained from Stokes flow numerical simulations, displayed in terms of the velocity magnitude $v$ normalized by its mean value $\bar{v}$, for (a) $S_{\mathrm{w}}=1.00$ and (b) $S_{\mathrm{w}}=0.71$. The colorbar is common to subfigures (a) and (b), with red colors indicating high velocities and blue colors low velocities. Regions where $\log_{10}(v/\bar{v})\leq-1$ are shown in the darkest blue tone. The solid phase (circular obstacles) is shown in gray and air clusters in black. (c) Partition of the velocity field for $S_{\mathrm{w}}=0.71$ into two types of flow structures: (i) backbone or preferential paths, depicted in red, and (ii) dead-end regions or stagnation zones of low velocity, depicted in blue. The number and size of the dead-ends increase with lower saturation. The inset depicts the geometry of a dead-end region, with $\ell$ representing the dead-end region's depth, i.e., extension from the contact with the backbone to the far liquid--gas boundary.
}
\end{figure*}

\begin{figure}
\centering
	\includegraphics{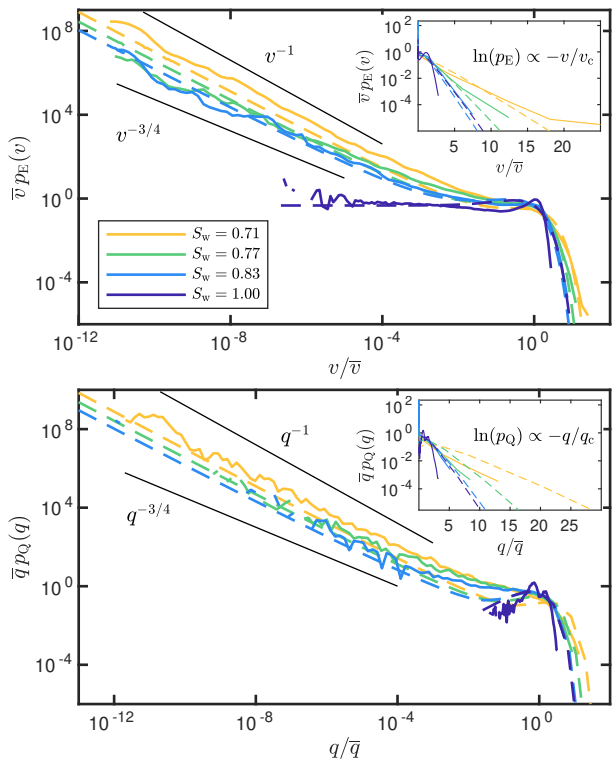}%
	\caption{\label{fig:velpdf} (color online). Numerical (continuous lines) and predicted (dashed lines) PDFs for (a) Eulerian velocities and (b) pore flow rates, normalized by their respective average values $\bar{v}$ and $\bar{q}$, for the four saturation degrees $S_{\mathrm{w}}=$ 1.00, 0.83, 0.77, and 0.71. The log-log scale highlights the scaling of low magnitudes; the power law scalings are shown for visual reference. Semi-log insets highlight the exponential behavior at high magnitudes.
}
\end{figure}

While the simulated velocity fields exhibit limited variability under saturated conditions (Fig.~\ref{fig:velfield}a), the flow heterogeneity is strongly enhanced in the unsaturated case (Fig.~\ref{fig:velfield}b). The introduction of air induces a partition of the flow field into two communicating flow structures~\cite{DeGennes1983}: a backbone of preferential flow paths, and dead-end regions (sometimes referred to as stagnation zones, although velocity is non-zero~\cite{Jimenez-Martinez2017,Jimenez-martinez2015}) that branch out from the backbone (Fig.~\ref{fig:velfield}c). This fundamental reorganization of flow compared to the saturated case leads to a marked change in the Eulerian velocity PDF $p_{\mathrm{E}}(v)$, with a strong increase of the probability of low velocities (Fig.~\ref{fig:velpdf}a). For all unsaturated cases, $p_{\mathrm{E}}(v)$ of low velocities is orders of magnitude larger than for the saturated system, as they are not only encountered close to the solid--liquid interfaces but also within dead-end regions (Fig.~\ref{fig:velfield}b). In the low-velocity range, $p_{\mathrm{E}}(v)$ undergoes a sharp transition from a plateau for $S_{\mathrm{w}}=1.00$ to a power-law-like behavior for $S_{\mathrm{w}}<1.00$. High velocities follow an exponential trend characterized by a saturation-dependent characteristic velocity $v_{\mathrm{c}}$.

We partition the flow field into backbone and dead-end regions (Fig.~\ref{fig:velfield}c) by selecting a velocity threshold at the transition between the power-law and exponential velocity regimes. Given the large velocity gradient at the boundaries between backbone and dead-ends, the regions distribution do not change significantly when varying this threshold. Results suggest a more accentuated flow separation with lower saturation, with dead-end regions increasing both in size and number as $S_{\mathrm{w}}$ decreases~\cite{SupplementaryMaterial}. The dead-end area PDF $p_\textrm{A}$ decays as a power-law whose variation in scaling reflects the increase in dead-end areas as saturation decreases~\cite{SupplementaryMaterial}. Note that previous experimental studies in 2D porous media have analyzed air cluster area distributions, rather than fluid dead-end area distributions, and found a power-law behavior with an exponential cutoff at large cluster sizes~\cite{Jimenez-Martinez2017,Tallakstad2009}.

To derive a theoretical framework for $p_{\mathrm{E}}(v)$, we first consider the local flow rate through a pore throat, or pore flow rate $q$. It is computed by integrating the pore-scale flow velocities over the cross section of the pore throat. For all unsaturated conditions, the PDF of pore flow rates over the ensemble of throats $p_{\mathrm{Q}}(q)$ shows a scaling similar to that of $p_{\mathrm{E}}(v)$ for both low and high magnitudes (Fig.~\ref{fig:velpdf}a). However, for $S_{\mathrm{w}}=1.00$, $p_{\mathrm{Q}}(q)$ increases with $q$ at low values instead of the plateau observed for $p_{\mathrm{E}}(v)$. In the saturated medium, $p_{\mathrm{Q}}(q)$ is well captured by the flow rate PDF in the backbone $p^{b}_{\mathrm{Q}}$, which follows a gamma distribution,
\begin{equation}
\label{eq:q_backbone}
p^{b}_{\mathrm{Q}}(q) = \frac{qe^{-q/q_{\mathrm{c}}}}{q^{2}_{\mathrm{c}}},
\end{equation}
where the saturation-dependent characteristic flow rate $q_{\mathrm{c}}$ controls the exponential high-flow tailing. This is consistent with the random aggregation model of~\cite{Alim2017}, based on the random splitting and merging of flow throughout the pore network~\cite{Coppersmith1996}. 

To model $p^{b}_{\mathrm{Q}}$ and $p_{\mathrm{E}}(v)$ for $S_{\mathrm{w}}<1.00$, we separately quantify the flow statistics in backbone ($p^{b}_{\mathrm{Q}}$) and dead-end ($p^{d}_{\mathrm{Q}}$) regions. We first determine the ratio $f$ of the area occupied by dead-end regions to the total medium area \cite{SupplementaryMaterial}, which ranges from $0.0072$ for $S_{\mathrm{w}}=0.83$ to $0.2601$ for $S_{\mathrm{w}}=0.71$. $p_{\mathrm{Q}}(q)$ is then related to $p^{b}_{\mathrm{Q}}$ and $p^{d}_{\mathrm{Q}}$ through
\begin{equation}
\label{eq:linearcomb}
p_{\mathrm{Q}}(q) = fp^{d}_{\mathrm{Q}}(q)+(1-f)p^{b}_{\mathrm{Q}}(q).
\end{equation}


Next, we determine $p^{d}_{\mathrm{Q}}$. Simulation data suggest that flow-rate magnitudes within dead-end regions decay exponentially with depth $0\leqslant z\leqslant\ell$~\cite{SupplementaryMaterial}, up to the total depth $\ell$ of the dead-end region, which extends from the contact with the backbone to the far liquid--gas boundary (see inset in Fig.~\ref{fig:velfield}c). Within each dead-end region, flow is driven by the pressure gradient across its connection to the backbone, which has a characteristic width on the order of the pore-throat size. Such exponential decay is consistent with the fundamental solutions of the Laplace equation for the propagation of a pressure perturbation applied on the boundary of a 2D domain~\cite{Bland1965}. We expect macroscopic pressure gradients within dead-ends regions to obey a Laplace equation resulting from Darcy's law~\cite{whitaker1986flow}. We thus approximate the flow rate decay along the depth as
\begin{equation}
\label{eq:qvariation}
q_\mathrm{d}(z|\ell)\approx q_{\mathrm{0}}e^{-z/a_{\mathrm{m}}}H(\ell-z),
\end{equation}
where $q_{\mathrm{0}}=q_\mathrm{d}(0|\ell)$ is the flow rate at the contact with the backbone and $H$ is the Heaviside step function. Taking $q_{\mathrm{0}}$ to be distributed according to Eq.~\eqref{eq:q_backbone} for $p^{b}_{\mathrm{Q}}$, we can now express the flow rate PDF in dead-ends given depth $\ell$ as 
\begin{equation}
\label{eq:PdfQsgivenL}
p^{d}_{\mathrm{Q}}(q|\ell)=\int_{0}^{\infty}dq_{\mathrm{0}} \,p^{d}_{\mathrm{Q}}(q|\ell,q_{\mathrm{0}}) \,p^{b}_{\mathrm{Q}}(q_{\mathrm{0}}),
\end{equation}
where $p^{d}_{\mathrm{Q}}(q|\ell,q_{\mathrm{0}})$ corresponds to an intra-region flow rate PDF for a given depth $\ell$ and a given local flow rate $q_{\mathrm{0}}$.

Setting $\ell \approx \sqrt{A}$, with $A$ the dead-end area, and averaging over areas, we obtain $p^{d}_{\mathrm{Q}}(q)$ ~\cite{SupplementaryMaterial}
\begin{equation}
\label{eq:PdfQsIntegral}
p^{d}_{\mathrm{Q}}(q)\approx \int_{0}^{\infty}dA \, p^{d}_{\mathrm{Q}}(q|\sqrt{A}) \, p_{\mathrm{A}}(A).
\end{equation}

We approximate the dead-end area PDF $p_{\mathrm{A}}$ by a Pareto distribution~\cite{SupplementaryMaterial}, 
\begin{equation}
\label{eq:Pareto}
p_{\mathrm{A}}(A)=\frac{\gamma}{a^{2}_{\mathrm{m}}}\left(\frac{A}{a^{2}_{\mathrm{m}}}\right)^{-1-\gamma} H(A-a^{2}_{\mathrm{m}}),
\end{equation}
where the exponent $\gamma$ decreases with decreasing $S_\mathrm{w}$, indicating broader dead-end area variability. We consider the minimum area of a dead-end region to be equal to the area of one pore throat, approximated as $a_{\mathrm{m}}^2$. 

Using the expressions for $p^{b}_{\mathrm{Q}}(q)$ (Eq.~\eqref{eq:q_backbone}) and $p^{d}_{\mathrm{Q}}(q)$ (Eq.~\eqref{eq:PdfQsIntegral}) in Eq.~\eqref{eq:linearcomb}, we deduce an expression for $p_{\mathrm{Q}}(q)$~\cite{SupplementaryMaterial}. In the low-flow regime, $q\ll q_{\mathrm{c}}$, the latter is controlled by the dead-end contribution as long as $f\neq0$, corresponding to $S_{\mathrm{w}}<1$. Taylor expansion of Eq.~\eqref{eq:PdfQsgivenL} for $q\ll q_{\mathrm{c}}$ together with Eq.~\eqref{eq:PdfQsIntegral} leads to
\begin{equation}
\label{eq:PdfQlow}
p_{\mathrm{Q}}(q)\approx \frac{2\gamma f}{q(1+2\gamma)}\left[\ln \left(\frac{q_{\mathrm{c}}}{q}\right)\right]^{-1-2\gamma}.
\end{equation}

Thus, our model predicts that, under unsaturated conditions, $p_{\mathrm{Q}}(q)$ scales for low flow rates as a power law, $q^{-1}$, corrected by a logarithmic factor. This factor is raised to a power controlled by $p_{\mathrm{A}}(A)$ through the exponent $\gamma$. In the particular case $f=0$, corresponding to $S_\mathrm{w}=1$, Taylor expansion of $p^{b}_{\mathrm{Q}}$ (Eq.~\eqref{eq:q_backbone}) for low $q$ leads to $p_{\mathrm{Q}}(q)\approx q/q_c^2$, linear in $q$. Proceeding similarly for $q\gtrsim q_{\mathrm{c}}$, for which we must consider contributions from both $p^{b}_{\mathrm{Q}}(q)$ and $p^{d}_{\mathrm{Q}}(q)$, we obtain the exponential decay
\begin{equation}
\label{eq:PdfQhigh}
p_{\mathrm{Q}}(q)\approx \left[\frac{2\gamma f}{1+2\gamma}+(1-f)\frac{q}{q_{\mathrm{c}}}\right]\frac{e^{-q/q_{\mathrm{c}}}}{q_{\mathrm{c}}}.
\end{equation}

We now turn to $p_{\mathrm{E}}(v)$, characterizing velocity variability across the medium. It results from the combined effect of $p_\mathrm{Q}(q)$ and the intra-throat variability arising from the local velocity profile within each throat. Thus, these two PDFs are related by
\begin{equation}
\label{eq:definitionPE}
p_\mathrm{E}(v) = \int_0^{\infty} dq \, \,  p_\mathrm{Q}(q) p_\mathrm{E}(v | q),
\end{equation}
where $p_\mathrm{E}(\cdot | q)$ is the PDF of velocities associated with a pore throat characterized by the flow rate $q$. Since pore throat widths are comparable in size to the channel thickness $h$, we take into consideration the impact of the confinement by the top and bottom plaque in the third dimension on the intra-pore, depth-averaged 2D velocity profile. The latter differs from the parabolic profile expected in a purely-2D scenario~\cite{SupplementaryMaterial}. Given the low variability of pore-throat sizes across the medium, we approximate throat widths by their average value $a_{\mathrm{m}}$. The Eulerian velocity PDF associated with the velocity profile across a pore throat with a corresponding local flow rate $q$ is then~\cite{SupplementaryMaterial}
\begin{equation}
\label{eq:pE_v_givenq}
p_\mathrm{E}(v | q) = \frac{2h}{\pi a_{\mathrm{m}} v_{\mathrm{max}}(q)}\frac{(C-1)H[v_{\mathrm{max}}(q)-v]}{\sqrt{[C-(C-1)v/v_{\mathrm{max}}(q)]^2-1}},
\end{equation}
where $C=\cosh(\pi a_{\mathrm{m}}/2h)$, and $v_{\mathrm{max}}(q)=\alpha q/(h a_\mathrm{m})$ is the maximum velocity within the pore throat, with
\begin{equation}
	\label{eq:ParameterAlpha}
	\alpha = 2\left(1+\mathrm{coth}\left(\frac{\pi a_{\mathrm{m}}}{4h}\right)\left[\mathrm{coth}\left(\frac{\pi a_{\mathrm{m}}}{4h}\right)-\frac{4h}{\pi a_{\mathrm{m}}}\right]\right)^{-1}.
\end{equation}

Under saturated conditions, the integral in Eq.~\eqref{eq:definitionPE} can then be 
approximated in the ranges $v\ll v_{\mathrm{c}}$ and $v\gtrsim v_{\mathrm{c}}= q_{\mathrm{c}}/ (a_\mathrm{m} h)$, respectively, by using Eq.~\eqref{eq:q_backbone}, as~\cite{SupplementaryMaterial}
\begin{gather}
\label{eq:PESaturated_low}
p_{\mathrm{E}}(v)\approx \frac{2h}{\pi a_{\mathrm{m}} \alpha v_{\mathrm{c}}}\mathrm{tanh}\left(\frac{\pi a_{\mathrm{m}}}{4h}\right),
\\
\label{eq:PESaturated_high}
\text{and ~ }p_{\mathrm{E}}(v)\approx \frac{2h}{\alpha a_{\mathrm{m}}v_{\mathrm{c}}}\sinh\left(\frac{\pi a_{\mathrm{m}}}{4h}\right)\sqrt{\frac{v}{\pi \alpha v_{\mathrm{c}}}}e^{-\frac{v}{\alpha v_{\mathrm{c}}}}.
\end{gather}
Equation~\eqref{eq:PESaturated_low} describes a low-velocity plateau, while Eq.~\eqref{eq:PESaturated_high} encodes exponential tailing at large velocities.  

Under unsaturated conditions, the previous derivation holds for the backbone component. Similar to the situation regarding the pore flow rates, $p_{\mathrm{E}}(v)$ for low velocities is dominated by the dead-end regions, while for $v\gtrsim v_{\mathrm{c}}$ the contribution of both backbone and dead-ends matters. The low-velocity behavior is controlled by low flow rates. For small $q$,  $p_\mathrm{E}(v | q)$ becomes arbitrarily narrow, because the maximum velocity is linear in $q$, see Eq.~\eqref{eq:pE_v_givenq}. Accordingly, $p_{\mathrm{E}}(v)$ is well approximated for low $v$ by setting $p_{\mathrm{E}}(v|q)\approx \delta[v-q/(h a_{\mathrm{m}})]$ in Eq.~\eqref{eq:definitionPE}, where $\delta(\cdot)$ is the Dirac delta, and using Eq.~\eqref{eq:PdfQlow} for $p_\mathrm{Q}(q)$. We obtain, for the range $v\ll v_{c}$,
\begin{equation}
\label{eq:PdfElow}
p_{\mathrm{E}}(v)\approx \frac{2\gamma f}{v(1+2\gamma)}\left[\ln\left(\frac{v_{\mathrm{c}}}{v}\right)\right]^{-1-2\gamma}.
\end{equation}

Analogously, for the range $v\gtrsim v_{\mathrm{c}}$, $p_{\mathrm{E}}(v)$ can be computed using Eq.~\eqref{eq:PdfQhigh} for $p_\mathrm{Q}(q)$, and Taylor expanding Eq.~\eqref{eq:pE_v_givenq} for high pore velocities $v\approx v_\textrm{max}(q)$, which leads to
\begin{equation}
\label{eq:PdfEhigh}
p_{\mathrm{E}}(v)\approx \left[\frac{2\gamma f}{1+2\gamma}\sqrt{\frac{\alpha v_{\mathrm{c}}}{v}}+(1-f)\sqrt{\frac{v}{\alpha v_{\mathrm{c}}}}\right]\frac{C_* e^{-\frac{v}{\alpha v_{\mathrm{c}}}}}{\alpha v_{\mathrm{c}}},
\end{equation}
where $C_*=2h\sinh\left(\pi a_{\mathrm{m}}/4h\right)/(\sqrt{\pi}a_{\mathrm{m}})$. Note that for large $h$ values compared to $a_\text{m}$, Eqs.~\eqref{eq:PESaturated_low}--\eqref{eq:PdfEhigh} reduce to expressions that correspond to those obtained under the assumption of a Poiseuille velocity profile (fully-2D case)~\cite{SupplementaryMaterial}.

Figure~\ref{fig:velpdf} shows the predictions (dashed lines) for both $p_\mathrm{Q}(q)$ and $p_{\mathrm{E}}(v)$. The model successfully captures the different regimes and scaling variation for the various $S_{\mathrm{w}}$. The low-velocity plateau for $S_{\mathrm{w}}=1.00$ is also captured. The results shown here correspond to numerical computation of the full theoretical PDFs according to Eqs.~\eqref{eq:q_backbone}, \eqref{eq:linearcomb}, \eqref{eq:PdfQsIntegral}, and~\eqref{eq:definitionPE}. Further details on the regime scalings and parameter values can be found in~\cite{SupplementaryMaterial}.

\begin{figure}
\centering
	\includegraphics{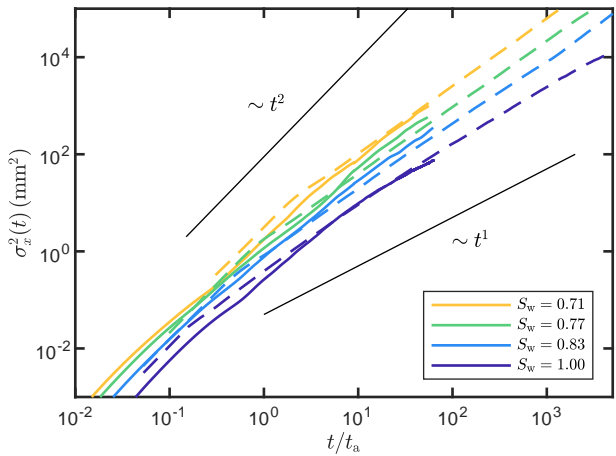}%
	\caption{\label{fig:ctrw} (color online). Advective dispersion $\sigma_{x}^2$ in time for all four saturation degrees $S_{\mathrm{w}}=$ 1.00, 0.83, 0.77, and 0.71. Time is normalized by the advective time $t_\mathrm{a}=\lambda/\bar{v}$ over the mean pore size $\lambda$. The plot compares $\sigma_{x}^2$ from the particle tracking analysis (continuous lines) with $\sigma_{x}^2$ from a CTRW approach computed using the predicted velocity PDF $p_{\mathrm{E}}(v)$ (dashed lines). Scalings for a ballistic $(\sigma_{x}^2\sim t^2)$ and a Fickian $(\sigma_{x}^2\sim t^1)$ regime are also displayed for reference. Time has been normalized by the corresponding advective time $t_{\mathrm{a}}$ of each saturation.}
\end{figure}

To investigate the consequences of these broad velocity distributions for transport, we compute dispersion through advective particle tracking simulations. Details on these simulations are presented in~\cite{SupplementaryMaterial}. Particle positions are tracked isochronically over fixed time steps $\Delta t$ ($t$-Lagrangian sampling). We compute the time evolution of the variance $\sigma_x^2$ of longitudinal particle positions. Lower saturation induces larger particle dispersion due to the increased velocity heterogeneity, as discussed above. At early times, a ballistic regime, $\sigma_x^2 \sim t^2$ is observed in Fig.~\ref{fig:ctrw}, for all $S_{\mathrm{w}}$, and then transitions to an asymptotic superdiffusive regime. The crossover time between the ballistic and asymptotic regimes is also larger for smaller $S_{\mathrm{w}}$, i.e., the Lagrangian correlation length $\zeta_x$ of velocities along the mean flow direction increases with decreasing saturation.

To develop a transport modeling framework that links the dispersion dynamics to hydrodynamics, we employ a CTRW approach~\cite{Berkowitz2006,Dentz2016,Cortis2004}. The CTRW framework used here models transport in terms of Lagrangian particles taking fixed spatial steps of length $\zeta_x$ along the mean flow direction ($s$-Lagrangian sampling). Particle velocities remain constant over a step and are assumed to fully decorrelate between steps. The velocity variability in the underlying medium is treated statistically, and the particle velocity in each step is sampled independently from the equilibrium $s$-Lagrangian velocity distribution, which is given by the flux-weighted Eulerian PDF, $p_\mathrm{s}(v)= vp_E(v)/\bar v$~\cite{Dentz2016}. This approach captures the intermittent nature of the $t$-Lagrangian velocity signal through the distributed waiting times to cross the fixed distance $\zeta_x$, arising from velocity variability sampled along streamlines~\cite{SupplementaryMaterial}.

To assess the applicability of our theoretical model to predict advective transport, we employ $p_\text{s}(v)$ defined from the predicted $p_{\mathrm{E}}(v)$ (dashed lines in Fig.~\ref{fig:velpdf}a) in the CTRW description. Fig.~\ref{fig:ctrw} shows the dispersion $\sigma_x^2$ computed from the resulting CTRW for each $S_{\mathrm{w}}$ (dashed lines), together with $\sigma_x^2$ computed from the fully-resolved particle tracking simulations. Dispersion is well predicted over both the ballistic and superdiffusive regimes, and so the impact of $S_{\mathrm{w}}$ on the temporal scaling. A slight overestimation of early-time dispersion is visible for $S_{\mathrm{w}}=1.00$, which might be explained by the assumption of full velocity decorrelation beyond the correlation length $\zeta_x$. This discrepancy is less pronounced for $S_{\mathrm{w}}<1.00$. Late-time dispersion is well captured in all cases, exhibiting more pronounced superdiffusive behavior for the unsaturated cases. Overall, these results support the suitability of both our theoretical description of velocity statistics and the CTRW to predict advective transport in unsaturated porous media. This conclusion is also supported by similar findings obtained when comparing these results to a CTRW parameterized according to $p_{\mathrm{E}}(v)$ computed from the flow simulations~\cite{SupplementaryMaterial}.

The CTRW provides an efficient computational approach that permits calculations across time scales that may be prohibitively expensive for more traditional numerical approaches such as standard particle tracking based on the pore-scale velocity field. Furthermore, it provides a theoretical framework to quantify and understand the relationship between dispersive scalings and velocity variability. In particular, the late-time scalings are controlled by the low-velocity behavior of $p_{\mathrm{E}}(v)$. If $p_{\mathrm{E}}(v)$ exhibits power-law decay near $v=0$, $p_E(v)\sim v^{-\theta}$ with $0<\theta<1$, late-time dispersion scales like $\sigma^2_x\sim t^{1+\theta}$~\cite{Dentz2016}, between the Fickian and ballistic limits $\sigma^2_x\sim t$ and $\sigma^2_x\sim t^2$. The scalings found here for saturated and unsaturated conditions correspond to two contrasting edge-cases. Under saturated conditions, $\theta=0$ (Eq.~\eqref{eq:PESaturated_low}), which leads to logarithmically-enhanced Fickian dispersion~\cite{Dentz2016}. Note that pure power-law decay characterized by $\theta\geqslant1$ is not integrable near $v=0$, so this behavior cannot occur because it does not correspond to a well-defined $p_{\mathrm{E}}(v)$. In this sense, unsaturated conditions are characterized by maximal variability of low velocities, described by logarithmic corrections to power-law decay with $\theta=1$ (Eq.~\eqref{eq:PdfElow}). This leads, in contrast to the minimally-superdiffusive saturated case, to logarithmically-inhibited ballistic dispersion. In light of these considerations, along with the fact that the unsaturated $p_{\mathrm{E}}(v)$ is extremely broad (Fig.~\ref{fig:velpdf}), the apparent power-law scalings in Fig.~\ref{fig:ctrw} vary slowly with time, as logarithmic corrections and the effect of progressively lower velocities come into play. A rigorous derivation of asymptotic dispersion scalings is beyond the scope of this work and will be presented elsewhere.

Our findings uncover the role of flow architecture in unsaturated porous media in the emergence of anomalous transport dynamics. 
We have presented a new theoretical framework for the prediction of pore-scale flow PDFs and advective transport capturing the impact of liquid-phase saturation. Results reveal that the introduction of an immiscible gas phase leads to a shift in the scaling of the velocity PDFs that induces a sharp transition to strongly anomalous transport. Under saturated conditions, dispersion is quasi-Fickian. In contrast, even under slightly unsaturated conditions, dispersion becomes quasi-ballistic. In practice, this superdiffusive dispersion behavior is sustained until low-velocity cutoffs introduced by additional processes, such as diffusion, become relevant. The long-term residence time of a particle in a dead-end region is eventually controlled by molecular diffusion, effectively cutting off extreme slow velocities~\cite{DeGennes1983}. While in the presence of diffusion the transport is thus always asymptotically Fickian at sufficiently late times, the dispersive scalings related to the velocity variability remain relevant over significant time scales. Our CTRW formulation also opens the door to the quantification of nontrivial scalings of dispersion~\cite{bijeljic2006pore,aquino2021diffusing}.

The theoretical formulation developed here for flow and velocity PDFs depends only on a small set of parameters, which reflect characteristics of the porous medium ($a_{\mathrm{m}}$ and $h$), the spatial distribution of phases in the system ($\gamma$ and $f$), and flow properties ($\zeta_x$ and $\chi$, along with the flow tailing scale $q_{\mathrm{c}}$, used to determine $v_{\mathrm{c}}=q_c/(ha_{\mathrm{m}})$). The dead-end area distribution is assumed to follow a power-law, as suggested by the numerical data. We expect this structure to be robust for (quasi-)2D systems, independently of the detailed pore geometry. In particular it holds for fully 2D systems, and so does the spatial distribution of pore flow rates in dead-end regions (Eq.~\eqref{eq:qvariation}), so that our theory also provides valid predictions for fully 2D systems when considering the limit $h/a_\text{m} \gg 1$ \cite{SupplementaryMaterial}. Generalization to 3D porous media may be undertaken with a similar approach, although we expect the topology of the backbone and dead-end regions to be significantly altered in 3D. We have studied a relatively homogeneous porous medium, and an important open question is the effect of broader pore size variability~\cite{DeAnna2017} under unsaturated conditions. Furthermore, the upscaling of flow and transport presented here is a first step towards theoretical assessment of mixing and chemical reactions in unsaturated porous media, which are essential processes for the analysis and optimization of environmental and industrial systems.

\begin{acknowledgments}
A.V.-P. and J. J.-M. gratefully acknowledge the financial support from the Swiss National Science Foundation (SNF, grant Nr. 200021 178986). T.A. is supported by a Marie Sk\l odowska Curie Individual Fellowship, under the project \textit{ChemicalWalks} 792041.
Y.M. and T.L.B. acknowledge funding from ERC project \textit{ReactiveFronts} 648377. Numerical flow simulations were performed using the software COMSOL Multiphysics.
\end{acknowledgments}

\nocite{Bruus}
\bibliography{BibliographyPRL2020_20210314}

\end{document}


\title{Supplemental Material\\ Sharp transition to strongly anomalous transport in unsaturated porous media}

\author{Andr\'{e}s Vel\'{a}squez-Parra}
\affiliation{Department of Water Resources and Drinking Water, Swiss Federal Institute of Aquatic Science and Technology, Eawag, D\"{u}bendorf, Switzerland}
\affiliation{Department of Civil, Environmental and Geomatic Engineering, ETH Z\"{u}rich, Z\"{u}rich, Switzerland}

\author{Tom\'{a}s Aquino}
\affiliation{Univ. Rennes, CNRS, G\'{e}osciences Rennes, UMR 6118, 35000 Rennes, France}

\author{Matthias Willmann}
\affiliation{Department of Civil, Environmental and Geomatic Engineering, ETH Z\"{u}rich, Z\"{u}rich, Switzerland}

\author{Yves M\'{e}heust}
\affiliation{Univ. Rennes, CNRS, G\'{e}osciences Rennes, UMR 6118, 35000 Rennes, France}

\author{Tanguy Le Borgne}
\affiliation{Univ. Rennes, CNRS, G\'{e}osciences Rennes, UMR 6118, 35000 Rennes, France}

\author{Joaqu\'{i}n Jim\'{e}nez-Mart\'{i}nez}
\email[]{joaquin.jimenez@eawag.ch / jjimenez@ethz.ch}
\affiliation{Department of Water Resources and Drinking Water, Swiss Federal Institute of Aquatic Science and Technology, Eawag, D\"{u}bendorf, Switzerland}
\affiliation{Department of Civil, Environmental and Geomatic Engineering, ETH Z\"{u}rich, Z\"{u}rich, Switzerland}

\maketitle

In this Supplemental Material, we present additional details and derivations regarding the results reported in the main document. The Supplemental Material is organized in three appendices. Appendix~\ref{sec:FlowFieldsStructure} concerns additional results on the numerical velocity analysis for all analyzed saturation degrees. Further details on the derivation of the theoretical model for the prediction of the probability density function (PDF) of both velocities and flow rates through the pore throats are presented in~Appendix~\ref{sec:Model}. Finally, details on the performance and application of the model for the prediction of advective transport and dispersion dynamics are discussed in Appendix~\ref{sec:CTRW}.

\section{Porous medium structure and flow}
\label{sec:FlowFieldsStructure}

\begin{figure*}[]
\begin{center}
	\includegraphics{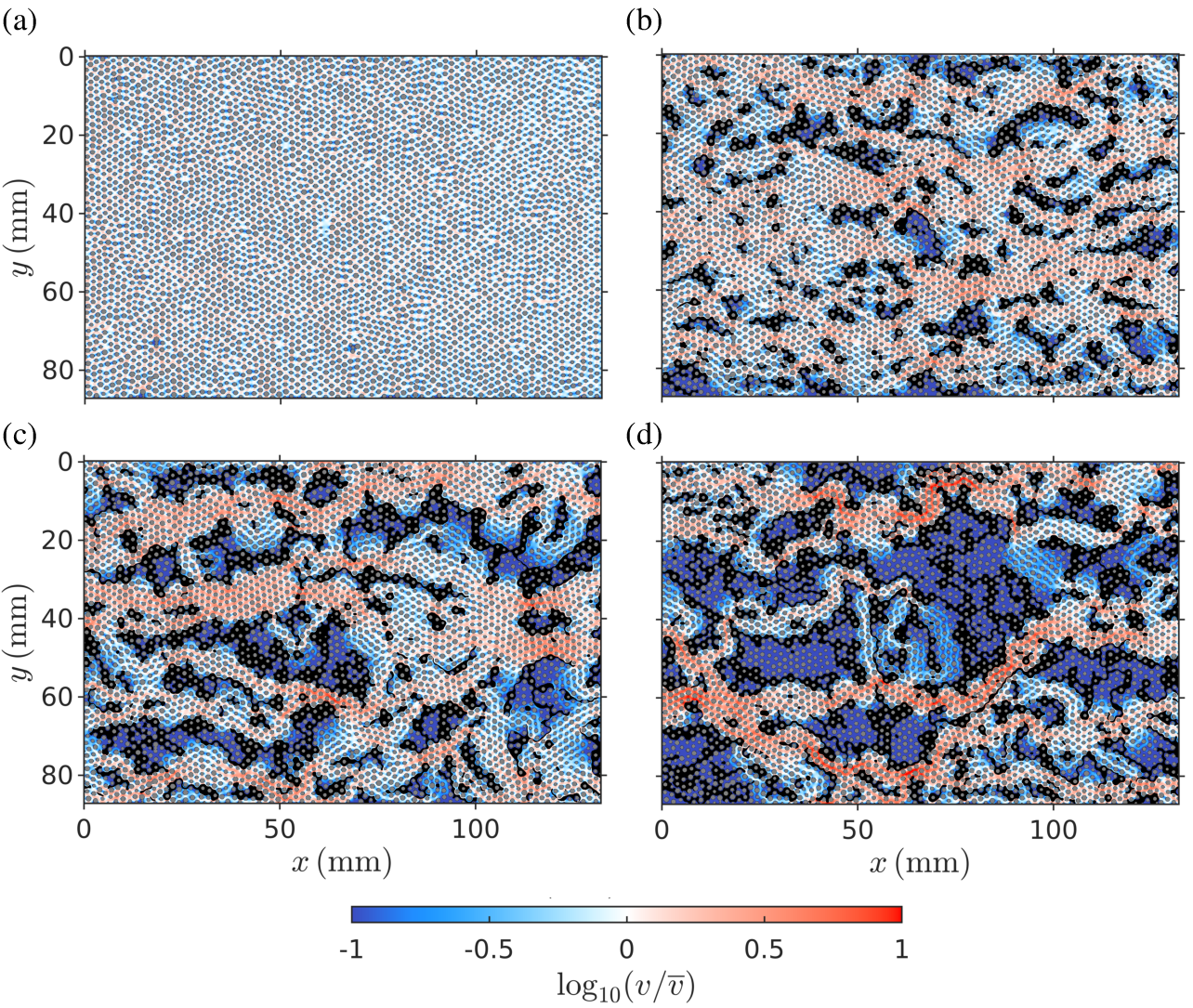}%
	\caption{\label{fig:velocityfields} Velocity fields obtained from Stokes flow numerical simulations, displayed in terms of the velocity magnitude $v$ normalized by its mean value $\bar{v}$, for (a) $S_{\mathrm{w}}=1.00$, (b) $S_{\mathrm{w}}=0.83$, (c) $S_{\mathrm{w}}=0.77$, and (d) $S_{\mathrm{w}}=0.71$. The color scale is common to all panels, with red colors indicating high velocities and blue colors low velocities. Regions where $\log_{10}(v/\bar{v})\leq-1$ are shown in the darkest blue tone. The solid phase (circular obstacles) is shown in gray and air clusters in black. More accentuated backbone and dead-end regions are induced with decreasing liquid phase saturation.}
	\end{center}
\end{figure*}

\begin{figure*}[]
\begin{center}
	\includegraphics[width=1\textwidth]{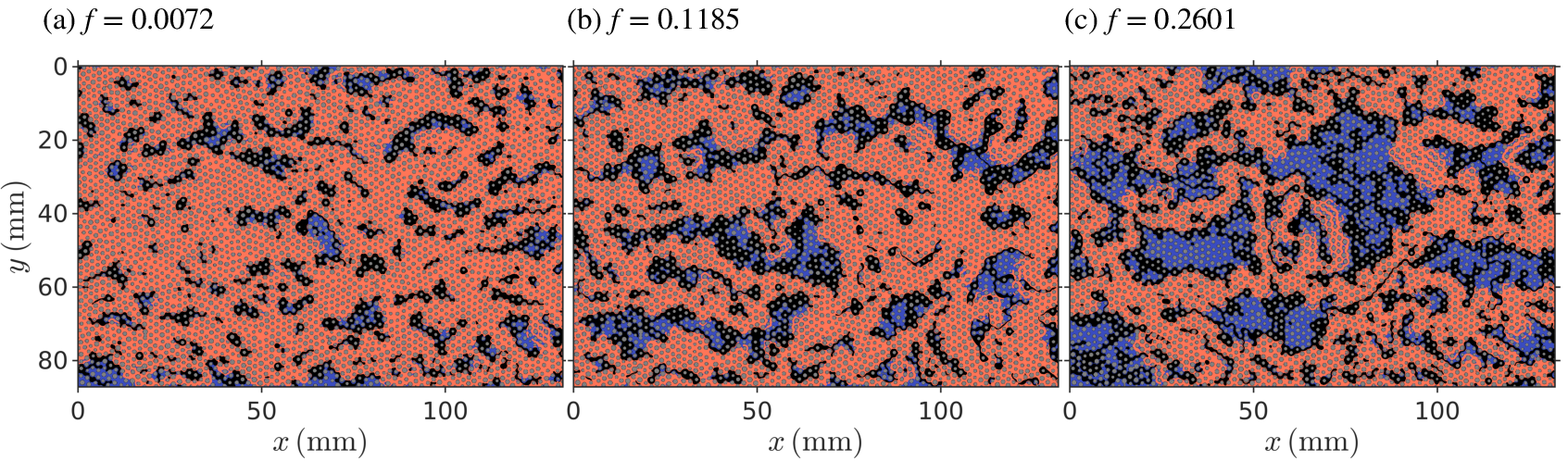}%
	\caption{\label{fig:doublestructure} Partition of the velocity field for (a) $S_{\mathrm{w}}=0.83$, (b) $S_{\mathrm{w}}=0.77$, and (c) $S_{\mathrm{w}}=0.71$ into two types of flow structures: (i) backbone or preferential paths of high velocity, depicted in red, and (ii) dead-end regions or stagnation zones of low velocity, depicted in blue. The number and size of dead-ends increase as liquid phase saturation decreases. The ratio $f$ of dead-end area to the total medium area is reported for each case.}
	\end{center}
\end{figure*}

Here, we present additional results on the flow fields obtained from the numerical Stokes flow flow simulations. The velocity fields for all analyzed saturation degrees are shown in Fig.~\ref{fig:velocityfields}. The associated partition of the three unsaturated flow fields into a backbone of preferential flow paths, and dead-ends, or regions of low velocity, for the unsaturated cases is shown in Fig.~\ref{fig:doublestructure}.

\begin{figure*}[]
	\includegraphics{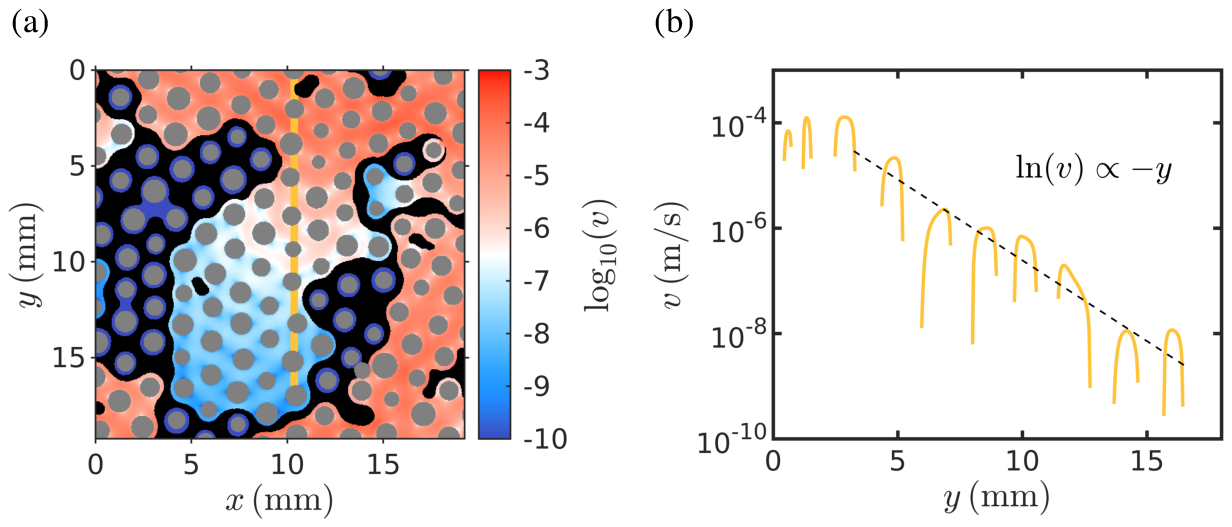}
	\caption{\label{fig:expVelDecay} (a) Close view of the velocity field for $S_{\mathrm{w}}=0.71$, highlighting the velocity field within a dead-end region. An observation cross-section is depicted as a continuous line. (b) Velocity variation over the depth of a dead-end region, here parallel to the $y$-axis, along the observation cross-section shown in (a). The velocity decays exponentially as a function of depth.
	}
\end{figure*}

\begin{figure*}[]
	\includegraphics{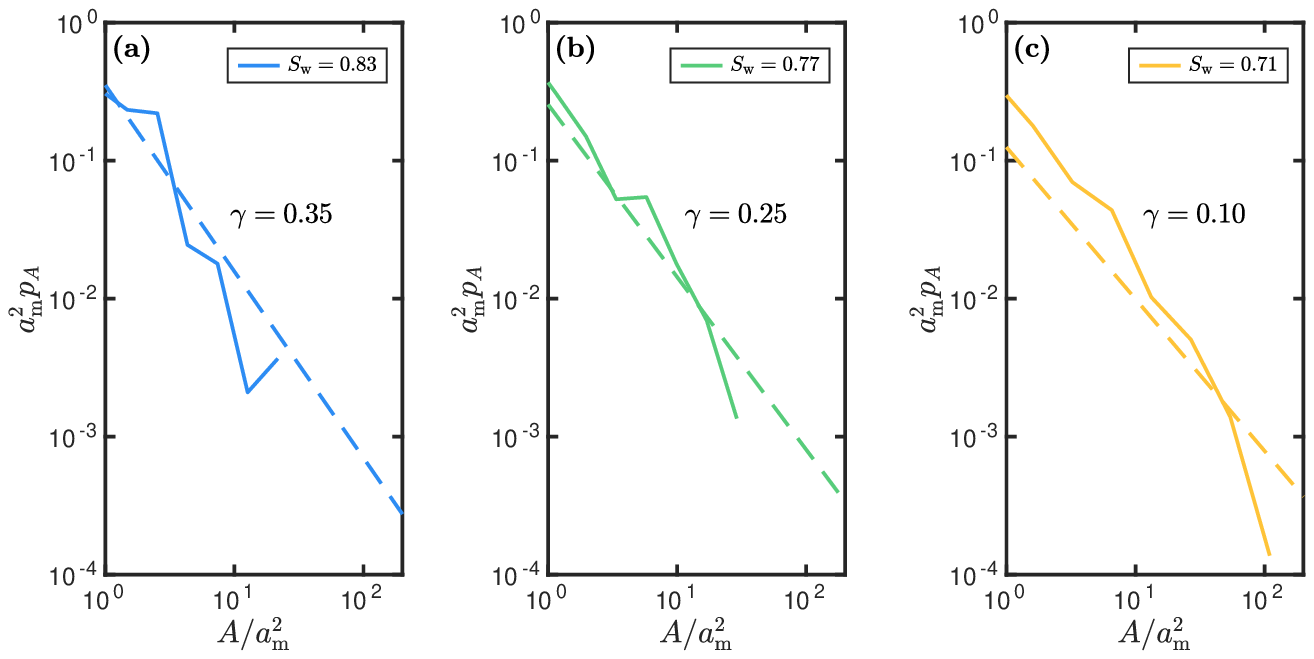}%
	\caption{\label{fig:paretofit} PDF of the dead-end areas $p_{A}$ for (a) $S_{\mathrm{w}}=0.83$, (b) $S_{\mathrm{w}}=0.77$, and (c) $S_{\mathrm{w}}=0.71$. For all three cases, $p_{A}$ is well approximated by a Pareto distribution (Eq. (6)) describing power-law decay. Corresponding values of the fitting parameter $\gamma$ describing a power law decay $\propto A^{-1-\gamma}$, which decreases with decreasing liquid phase saturation, are shown. Quantities are non-dimensionalized with respect to the area $a_{\mathrm{m}}^2$ associated with the average pore-throat aperture.}
\end{figure*}

We now turn our attention to the structure of dead-end regions in the three unsaturated flow cases. A representative example of the velocity variation occurring within a single dead-end region is presented in Fig.~\ref{fig:expVelDecay}. The PDF $p_A$ of dead-end areas, defined for a given flow field such that $p_A(A)\,dA$ is the probability of a uniformly randomly chosen dead-end region to have area in $[A,A+dA]$, is shown in Fig.~\ref{fig:paretofit} for each unsaturated flow field. The area PDFs were determined based on the flow field partitions of Fig.~\ref{fig:doublestructure}. As shown in the figure, the area PDFs are well approximated by a Pareto distribution describing power-law decay, see Eq. (6) of the main text.

\section{Theoretical model for flow and velocity PDFs}\label{sec:Model}

The assumptions and core of the approach behind our model for the flow and velocity PDFs are discussed in the main text. Here, we provide some additional details on the derivations of our main results.

First, consider the flow rate PDF in dead-end regions. As discussed in the main text, we approximate the decay of pore flow rate $q_\mathrm{d}(z|\ell)$ as a function of depth $z$ within a dead-end region of total depth $\ell$, as exponential, Eq. (3) (see also Fig.~\ref{fig:expVelDecay}). Since this flow rate profile is monotonically decreasing, the associated PDF for a given $q_0$ and $\ell$ can be computed as~\cite{aquino2021diffusing}
\begin{equation}
\label{eq:p_q_givenlandq0}
p^{d}_{\mathrm{Q}}(q|\ell,q_{\mathrm{0}})=\left(\ell \frac{dq_\mathrm{d}(z|\ell)}{dz}\Bigg|_{z=z_\textrm{q}(q)}\right)^{-1},
\end{equation}
where $z_\textrm{q}(q)$ is the point at which the flow has a given value $q$, i.e., $q_\mathrm{d}[z_\textrm{q}(q)|\ell]=q$. Thus, inverting Eq. (3) for depth as a function of flow rate, computing $dq_\mathrm{d}(z|\ell)/dz$, and substituting, Eq.~\eqref{eq:p_q_givenlandq0} becomes
\begin{equation}
\label{eq:p_q_given_lq0_solved}
p^{d}_{\mathrm{Q}}(q|\ell,q_{\mathrm{0}})=\frac{a_\mathrm{m}}{\ell q}H(q_0-q)H(q-q_0e^{-\ell/a_\mathrm{m}}),
\end{equation}
for the dead-end flow-rate PDF $p^{d}_{\mathrm{Q}}(\cdot|\ell,q_{\mathrm{0}})$, given maximum depth $\ell$ and flow rate $q_0=q_\mathrm{d}(0|\ell)$ at the entrance. Taking the flow rates $q_{0}$ at the entry of the dead-end region to be distributed according to the flow in the backbone (Eq. (1)) and averaging over the latter, we obtain Eq. (4). Using this result together with the Pareto area PDF, Eq. (6), we find an explicit form for the dead-end PDF given maximum depth $\ell$,
\begin{equation}
\label{eq:pqs_q_given_l}
p^{d}_{\mathrm{Q}}(q|\ell) = \frac{a_\mathrm{m}}{\ell q_{\mathrm{c}}q}\left[e^{-q/q_{\mathrm{c}}}(q+q_{\mathrm{c}})-\exp\left(-\frac{e^{\ell/a_{\mathrm{m}}}q}{q_{\mathrm{c}}}\right)\left(e^{\ell/a_{\mathrm{m}}}q+q_{\mathrm{c}}\right)\right].
\end{equation}

As discussed in the main text, we approximate $\ell \approx \sqrt{A}$ in terms of the corresponding dead-ends regions area, and we average over the dead-end areas to obtain the dead-end region flow PDF $p^{d}_{\mathrm{Q}}$, yielding Eq. (5). Expanding the integrand for $q\gtrsim q_{\mathrm{c}}$ using Eq.~\eqref{eq:pqs_q_given_l}, we obtain
\begin{equation}
\label{eq:pqs_qlargerqc}
p^{d}_{\mathrm{Q}}(q)\approx \int_{0}^{\infty}dA\,\frac{a_{\mathrm{m}}}{q_{\mathrm{c}}\sqrt{A}} e^{-q/q_{\mathrm{c}}} p_{\mathrm{A}}(A).
\end{equation}

Carrying out this integral analytically, and combining it with Eqs. (1) and (2), we obtain Eq. (7). For the PDF in the limit $q\ll q_{\mathrm{c}}$, notice that the nested exponential in Eq.~\eqref{eq:pqs_q_given_l} varies rapidly from zero to unity around $q=q_{\mathrm{c}} e^{-\ell / a_{\mathrm{m}}}$, so that it is well approximated by a cutoff for $A > [a_{\mathrm{m}}\mathrm{ln}(q_{\mathrm{c}}/q)]^2$. Again using Eq. (5), and further expanding for $q\ll q_{\mathrm{c}}$, this leads to
\begin{equation}
\label{eq:pqs_qsmallerqc}
p^{d}_{\mathrm{Q}}(q)  \approx \int\limits_{0}^{\infty}dA\,\frac{a_{\mathrm{m}}}{q_{\mathrm{c}}\sqrt{A}} \left[1-H \left(\left[a_{\mathrm{m}}\ln \left(\frac{q_{\mathrm{c}}}{q}\right)\right]^2 -A\right)\right]p_{\mathrm{A}}(A)=\int\limits_{[a_{\mathrm{m}}\mathrm{ln}(q_{\mathrm{c}}/q)]^2}^{\infty}\!\!\!\!\!\!\!dA\,\frac{a_{\mathrm{m}}p_{\mathrm{A}}(A)}{q_{\mathrm{c}}\sqrt{A}},
\end{equation}
where $H$ is the Heaviside step function. This integral can be computed analytically, and in combination with Eqs. (1) and (2), leads to Eq. (8).

The remainder of this appendix is concerned with the velocity PDFs. We first provide details on the full model, corresponding to our main results, which takes into account three-dimensional effects due to the finite medium thickness $h$. We then discuss its two-dimensional limit $a_\textrm{m}/h\to0$, in which the flow within pore throats reduces to a standard Poiseuille profile.

\subsection{Full formulation considering three-dimensional effects}\label{sec:Model3D}

\begin{figure*}[]
	\includegraphics[width=0.5\textwidth]{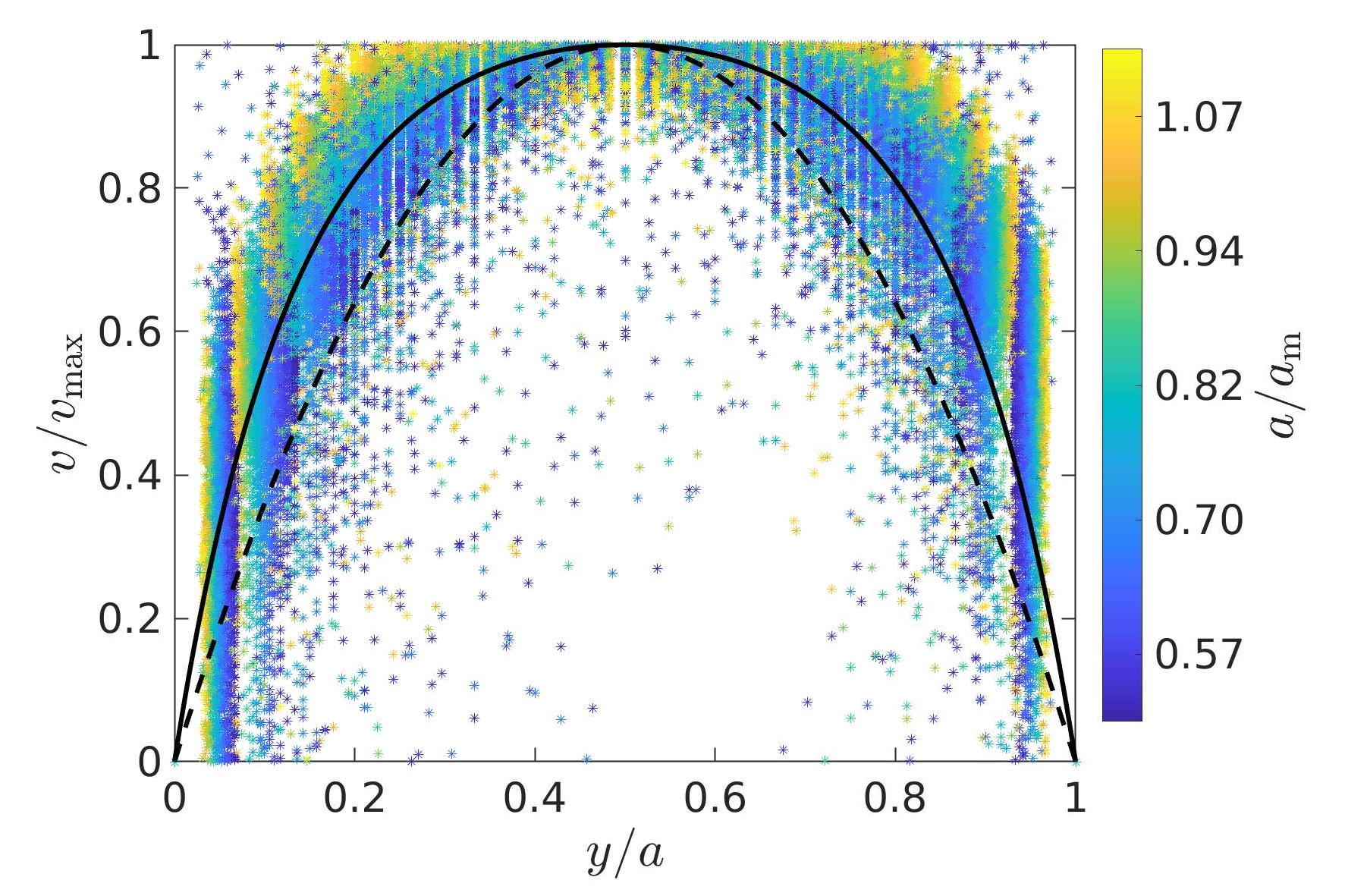}%
	\caption{\label{fig:velocityprofile} Spatial distribution of liquid-phase velocities $v$ within each pore throat in the medium along the direction $y$ transverse to the local main flow direction at each throat, under fully-saturated conditions ($S_{\mathrm{w}}=1.00$). Velocities are normalized by the maximum velocity $v_{\mathrm{max}}$ at each pore throat and $y$ coordinates are normalized by the corresponding pore throat size $a$. The plot highlights the variability of the velocity profile across pore throats. The color map represents the pore throat size, with dark blue tones corresponding to small pore throats and yellow colors representing large ones. The continuous line represents the mean velocity profile described by Eq.~\eqref{eq:Avgvelocityprofile}. The dashed line shows a parabolic (i.e., Poiseuille) velocity profile (Eq.~\eqref{eq:PoiseuilleVelProfile}), found only in the smaller pore throats.
	}
\end{figure*}

The porous medium used in the numerical Stokes flow simulations~\cite{Jimenez-Martinez2017} (see Fig.~\ref{fig:velocityfields}) has a pore throat width that is comparable in size to the channel thickness. Therefore, three-dimensional effects impact the intra-pore velocity profiles and the corresponding point statistics. Setting a local coordinate system at each pore throat, with $x$ representing the local mean flow direction, the throat width running parallel to $y$, and $z$ along the channel thickness, we analyze the resulting velocity profile over the $y$ direction, averaged over the ensemble of pore throats (see Fig.~\ref{fig:velocityprofile}).

We can express the velocity profile for Stokes flow over the channel thickness $h$ for a pore throat, approximated as a cuboid channel of width $a_\text{m}$ and thickness $h$, as a series in the form~\cite{Bruus}
\begin{equation}
\label{eq:Velocityprofile}
v_{\mathrm{x}}(y,z) = \frac{4h^2 \Delta p}{\pi^3\mu L} \sum_{n,\mathrm{odd}}^{\infty}\frac{1}{n^3}\left[1-\frac{\cosh(n\pi\frac{y}{h})}{\cosh(n\pi \frac{a_{\mathrm{m}}}{2h})}\right]\sin\left(n\pi \frac{z}{h}\right),
\end{equation}
where $\mu$ is the viscosity of the liquid (wetting) phase, $\Delta p$ is the pressure difference across the cuboid channel of length $L$, and the sum extends over odd values of $n$ as indicated. Averaging this function over $z$ values between $0$ and $h$, we obtain
\begin{equation}
\label{eq:Avgvelocityprofile}
\langle v_{\mathrm{x}}(y,z)\rangle_{z} = \frac{8h^2 \Delta p}{\pi^4\mu L} \sum_{n,\mathrm{odd}}^{\infty}\frac{1}{n^4}\left[1-\frac{\cosh(n\pi\frac{y}{h})}{\cosh(n\pi \frac{a_{\mathrm{m}}}{2h})}\right],
\end{equation}
where $\langle \cdot \rangle$ denotes the mean value. The velocity profile obtained with this expression is represented by the continuous line in Fig.~\ref{fig:velocityprofile}. For the dimensions of our porous medium ($h$ and $a_{\mathrm{m}}$), this series is governed by its first term, reducing the expression to
\begin{equation}
\label{eq:FirstTermVelocityprofile}
\langle v_{\mathrm{x}}(y,z)\rangle_{z} \approx \frac{h^2 \Delta p}{2\pi^4\mu L} \left[1-\frac{\cosh(2\pi\frac{y}{h})}{\cosh(\pi \frac{a_{\mathrm{m}}}{h})}\right].
\end{equation}

We are now in a position to derive a theoretical expression for the PDF of the depth-averaged velocity magnitudes in the $x$-$y$ plane. Normalizing the previous result by the maximum depth-averaged velocity $v_{\mathrm{max}}$ and then inverting for $y$, we obtain
\begin{equation}
\label{eq:VelProfileEq_Inverted}
y_\mathrm{v}(v) = \frac{h}{2\pi}\mathrm{arcosh}\left[C-\frac{v}{v_{\mathrm{max}}}\left(C-1\right)\right], \qquad C=\cosh\left(\frac{\pi a_{\mathrm{m}}}{2h}\right).
\end{equation}

Following the same procedure as for Eq.~\eqref{eq:p_q_given_lq0_solved}, we obtain Eq. (10) for the velocity PDF $p_\mathrm{E}(\cdot | q)$ given pore flow rate $q$. In order to characterize the maximum velocity in the pore  $v_{\mathrm{max}}(q)$ as a function of the flow rate $q$, we multiply Eq. (10) by velocity $v$ and integrate, which leads to the mean pore velocity
\begin{equation}
\label{eq:meanporevelocity}
v_{\mathrm{m}}(q)=\frac{v_{\mathrm{max}}(q)}{2}\left(1+\mathrm{coth}\left(\frac{\pi a_{\mathrm{m}}}{4h}\right)\left[\mathrm{coth}\left(\frac{\pi a_{\mathrm{m}}}{4h}\right)-\frac{4h}{\pi a_{\mathrm{m}}}\right]\right). 
\end{equation}

Relating the local flow rate $q$ to the local mean velocity by $q=ha_{\mathrm{m}}v_{\mathrm{m}}(q)$, we find $v_{\mathrm{max}}(q)=\alpha q/(h a_\mathrm{m})$, with $\alpha$ a constant given by Eq. (11).

As discussed in the main text, using these results and those for the flow rate PDF together with Eq. (9) allows us to obtain analytical scalings for the low- and high-velocity dependencies of the Eulerian velocity PDF $p_E$ by appropriate Taylor expansions. We note here that, for the range $v \gtrsim v_{\mathrm{c}}$, we expand the integrand for $v-\alpha q/(h a_{\mathrm{m}})\ll 1$ (velocities near the maximum pore velocity), which corresponds to a pole of the integrand and provides the dominant contribution in this case. We obtain, in terms of the Meijer $G$ function,
\begin{equation}
\label{eq:PEMeijerG3D}
p_{\mathrm{E}}(v) \approx \frac{\sqrt{2}\pi hv}{a_{\mathrm{m}}\alpha^2 v_{\mathrm{c}}^2} \left[ G^{20}_{24}\left(\frac{v^2}{4v_{\mathrm{c}}^2 \alpha^2}\middle|-\frac{1}{4},\frac{1}{4};-\frac{1}{2},\frac{1}{2},0,0\right) \\ -G^{20}_{24}\left(\frac{v^2}{4v_{\mathrm{c}}^2 \alpha^2}\middle|-\frac{1}{4},\frac{1}{4};0,0,-\frac{1}{2},\frac{1}{2}\right)\right].
\end{equation}

Expanding again for $v \gtrsim v_{\mathrm{c}}$, we obtain the high-velocity behavior for the saturated case, Eq. (13).
 
\subsection{Two-dimensional limit}\label{sec:Model2D}

In two-dimensional porous media, the velocity profile within pore throats is typically well approximated by a Poiseuille profile~\cite{DeAnna2017}. We now show that this assumption is equivalent to the previous formulation in the limit of large medium thickness or small pore throats, $a_{\mathrm{m}}/h \rightarrow0$, when the flow dynamics are effectively two-dimensional. As shown in Fig.~\ref{fig:velocityprofile}, the Poiseuille profile provides a good fit across the smaller pore throats in our medium, but, for wider throats, three-dimensional effects become relevant. Note that the arguments leading to the flow rate PDF make no use of the flow profile within pores, and therefore the $p_Q$ remains unchanged. The presence of three-dimensional effects impacts only the velocity PDF $p_E$, although, as we now show, the functional scalings at low and high velocity remain unchanged.

First, assume that the velocity field within a pore throat is well approximated by a Poiseuille profile. Then, for a local flow rate $q$ at the pore throat, the Eulerian velocity profile in the direction transverse to the flow is
\begin{equation}
\label{eq:PoiseuilleVelProfile}
v_{\mathrm{E}}(v|q)=\frac{3q}{2a_{\mathrm{m}}}\left(1-\frac{y^2}{a_{\mathrm{m}}^2}\right),
\end{equation}
where the mean velocity within the pore throat is given by $v_{\mathrm{m}}=ha_{\mathrm{m}}q$ and the maximum velocity is equal to $3v_{\mathrm{m}}/2$. Consider the case of fully-saturated conditions. The intra-pore PDF associated with the Poiseuille flow profile is~\cite{aquino2021diffusing}
\begin{equation}
\label{eq:PE_given_q_Poiseuille}
p_{\mathrm{E}}(v|q)=\frac{a_{\mathrm{m}} H[3q/(2a_{\mathrm{m}})-v]}{3q\sqrt{1-2a_{\mathrm{m}}v/(3q)}}.
\end{equation}
In the limit $a_{\mathrm{m}}/h \rightarrow 0$, we have $\alpha\to2/3$ (Eq. (11)), yielding the same result for the relationship between flow and maximum velocity mentioned previously for the full formulation. Substituting this in Eq. (10) and again taking the limit $a_{\mathrm{m}}/h \rightarrow 0$, we obtain Eq.~\eqref{eq:PE_given_q_Poiseuille}. This shows that intra-pore Poiseuille velocity profiles are a limit case of the general model derived in section~\ref{sec:Model3D}.

Based on the definition of the overall velocity PDF $p_{\mathrm{E}}(v)$, Eq. (9), and recalling that $p_{\mathrm{Q}}(q)$ is given by Eq. (1) for the backbone, the global PDF can then written in terms of the Meijer $G$ function as
\begin{equation}
\label{eq:PEPoisuille_MeijerG}
p_{\mathrm{E}}(v)=\frac{\sqrt{\pi}}{3v_{\mathrm{c}}}\frac{2v}{3v_{\mathrm{c}}}G^{20}_{12}\left(\frac{3v}{3v_{\mathrm{c}}}\middle|-\frac{1}{2}:-1,0\right),
\end{equation}
where the characteristic velocity $v_{\mathrm{c}}=q_{\mathrm{c}}/(h a_{\mathrm{m}})$. For small velocities, $v\ll v_{\mathrm{c}}$, this admits the series expansion
\begin{equation}
\label{eq:PE_Poiseuille_saturated_low}
p_{\mathrm{E}}(v)\approx \frac{1}{3v_{\mathrm{c}}}\left[1+\left(1-\ln \left[\frac{2e^{2\gamma_{\mathrm{E}}+\digamma(-1/2)}v}{3v_{\mathrm{c}}}\right]\right)\frac{v}{3v_{\mathrm{c}}}\right],
\end{equation}
where $\digamma$ is the digamma function, defined in terms of the Gamma function $\Gamma$ as $\digamma(x)=d\ln \Gamma(x)/dx$, and $\gamma_{\mathrm{E}}\approx0.5772$ is the Euler-Mascheroni constant. We have $\digamma(-1/2)\approx 0.03649$. In particular, $p_{\mathrm{E}}(v)\approx p_{\mathrm{E}}(0)=1/(3v_{\mathrm{c}})$ for sufficiently small $v$, which agrees with Eq. (12) when $a_{\mathrm{m}}/h \rightarrow0$. For large velocities, $v\gtrsim v_{\mathrm{c}}$ we find
\begin{equation}
\label{eq:PE_Poiseuille_saturated high}
p_{\mathrm{E}}(v)\approx \frac{1}{3v_{\mathrm{c}}}\sqrt{\frac{2\pi v}{3v_{\mathrm{c}}}}e^{-\frac{2v}{3v_{\mathrm{c}}}},
\end{equation}
which is again equivalent to the limit $a_{\mathrm{m}}/h \rightarrow0$ of Eq. (13). Thus, the high-velocity behavior remains exponential, and the low-velocity behavior remains flat.

Next, we examine unsaturated conditions. For low velocities, we can again approximate $p_{\mathrm{E}}(v|q)\approx\delta[v-q/(ha_{\mathrm{m}})]$. Therefore, Eq. (14) remains valid without change. As before, the contribution of dead-end regions to the high-velocity behavior is controlled by the high-flow-rate behavior. The backbone contribution is proportional to Eq.~\eqref{eq:PE_Poiseuille_saturated high}. Proceeding similarly to the three-dimensional formulation, and using Eq. (8) for the flow contribution of flow rates in dead-end regions for $q\gtrsim q_{\mathrm{c}}$, we find
\begin{equation}
\label{eq:PE_Poiseuille_unsat_high}
p_{\mathrm{E}}(v)\approx\left[\frac{\gamma f}{1+2\gamma}\sqrt{\frac{3v_{\mathrm{c}}}{v}}+(1-f)\sqrt{\frac{v}{3v_{\mathrm{c}}}}\right]\frac{\sqrt{2\pi}}{3v_{\mathrm{c}}}e^{-\frac{2v}{3v_{\mathrm{c}}}},
\end{equation}
which agrees with the limit $a_{\mathrm{m}}/h \rightarrow0$ of Eq. (15).

\subsection{Model parameters and regime scalings}
\label{sec:VelPDFs}

Table~\ref{tab:values_parameters} summarizes the parameter values employed in our model results for the velocity and flow rate PDFs, as well as for the CTRW transport model discussed in the main text and Appendix~\ref{sec:CTRW}. In all cases, the average pore-throat width $a_{\mathrm{m}}=1.17\,\textrm{mm}$, the medium thickness $h=0.5\,\mathrm{mm}$, and the mean pore size  $\lambda=1.85\,\mathrm{mm}$, as measured.

\begin{table}
	\caption{Model parameters. The parameters employed to model the flow and velocity PDFs are the ratio $f$ of dead-end areas to total area, the Pareto exponent $\gamma$ of the dead-end area PDF, and the characteristic pore flow rate $q_{\mathrm{c}}$. The characteristic velocity $v_{\mathrm{c}}=q_c/(h a_m)$ is also reported. The CTRW transport model requires the tortuosity $\chi$ and the longitudinal correlation length $\zeta_x$ of Lagrangian velocities.}
	\renewcommand{\arraystretch}{0.8}
	\setlength{\tabcolsep}{6pt}
	\centering
	\begin{tabular}{ccccccc}
		\toprule
		\textbf{$S_{\mathrm{w}}$} & \textbf{$f\,(-)$} & \textbf{$\gamma \,(-)$} & \textbf{$q_{\mathrm{c}}\,(\mathrm{m^3/s})$} & \textbf{$v_{\mathrm{c}}\,(\mathrm{m/s})$} & \textbf{$\chi \,(-)$} & \textbf{$\zeta_x \,(\lambda)$}\\
		\midrule
		$1.00$ & $0.000$ & $-$  & $1.19\times 10^{-11}$ & $2.04\times 10^{-5}$ & $1.10$ & $0.45$\\
		$0.83$ & $0.072$ & $0.35$ & $2.90\times 10^{-12}$ & $4.95\times 10^{-6}$ & $1.18$ & $1.10$\\
		$0.77$ & $0.119$ & $0.25$ & $4.82\times 10^{-12}$ & $8.25\times 10^{-6}$ & $1.23$ & $2.23$\\
		$0.71$ & $0.260$ & $0.10$ & $9.65\times 10^{-12}$ & $1.65\times 10^{-5}$  & $1.29$ & $6.70$\\
		\bottomrule
		\end{tabular}
	\label{tab:values_parameters}
\end{table}

\begin{figure*}[]
	\includegraphics{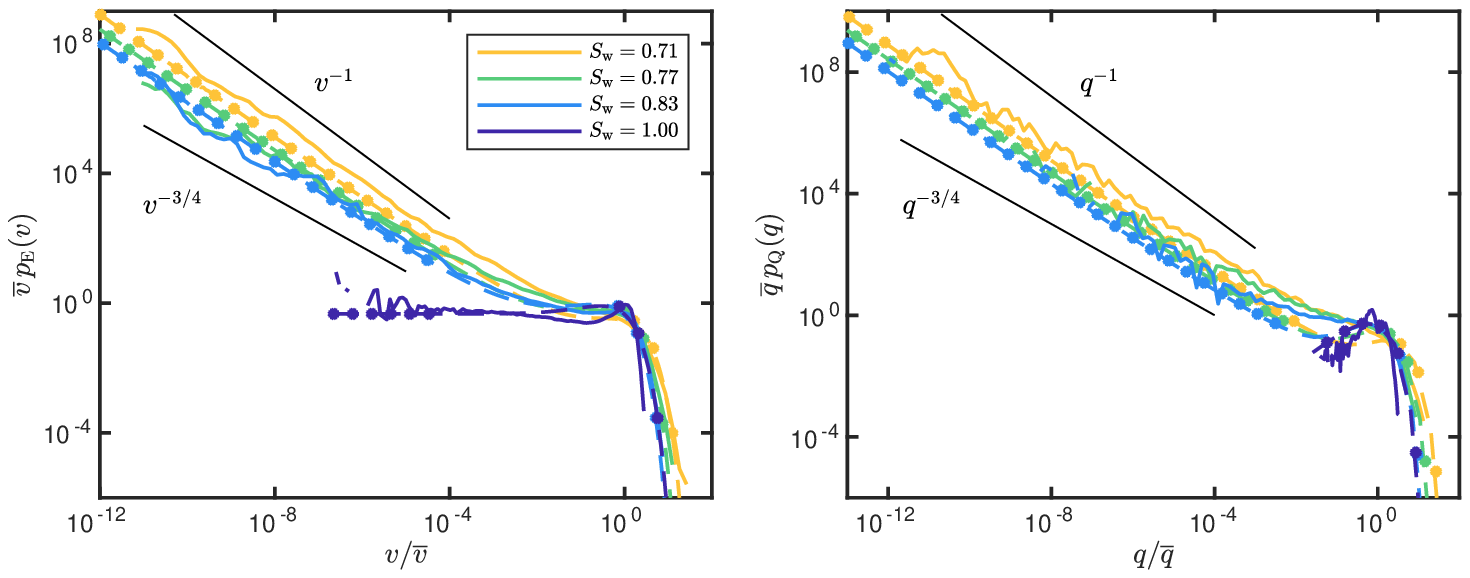}%
	\caption{\label{fig:q_v_PDF_IntegralModelData} (a) Eulerian velocity PDFs and (b) local flow rate PDFs for all analyzed saturation degrees $S_{\mathrm{w}}=1.00$, $0.83$, $0.77$, and $0.71$. Continuous lines represent the data from Stokes flow simulations; dashed lines represent full numerical computation of the integral representation of the theoretical PDFs discussed in the main text; and circular markers represent the analytical low- and high-magnitude regime scalings. Plots are non-dimensionalized according to the mean velocity $\overline v$ and the mean pore flow rate $\overline q$ of the numerical data set, respectively. 
}
\end{figure*}

As a verification of the low- and high-magnitude analytical scaling results, Figure~\ref{fig:q_v_PDF_IntegralModelData} shows comparison of the numerical data from Stokes flow simulations, the full theoretical PDFs computed numerically from the general theoretical expressions as discussed in the main text, and the regime scalings. Figure~\ref{fig:q_v_PDF_IntegralModelData}a concerns the velocity PDFs, with regime scalings given by Eqs. (12), (13), (14), and (15), whereas Fig.~\ref{fig:q_v_PDF_IntegralModelData}b shows the flow-rate PDFs, with scalings according to Eqs. (1), (7), and (8).

\section{Particle tracking simulations and the continuous time random walk model}\label{sec:CTRW}

Here, we discuss some details of the numerical particle tracking simulations used for direct numerical simulation of advective transport and the continuous time random walk model used to predict dispersion based on the theoretical predictions for the velocity PDF. The particle tracking simulation employs a flux-weighted injection of $10^4$ particles over an area close to the inlet of the porous medium along the inlet boundary, covering the full medium width in the $y$-direction and a length equal to the average grain size (i.e., $0.83\,\mathrm{mm}$) along the $x$-axis. Particle positions are tracked isochronically over fixed time steps $\Delta t$ ($t$-Lagrangian sampling). For $S_{\mathrm{w}}=1.00$, $\Delta t=0.05\,t_{\mathrm{a}}$, where $t_\mathrm{a}=\lambda/\bar{v}$ is the advective time over the mean pore size $\lambda$. For the unsaturated cases,  $\Delta t$ ranges between $0.021\,t_{\mathrm{a}}$ and $0.029\,t_{\mathrm{a}}$. A 100-times finer time discretization is introduced at early times to improve resolution in the ballistic regime.

To assess the applicability of the theoretical model introduced in section~\ref{sec:Model3D} for the prediction of advective transport, we employ a CTRW model. The CTRW description discussed in the main text corresponds to a set of stochastic recursion relations for the particle positions and times $X_k$ and $T_k$ at the completion of the $k$th step,
\begin{align}
X_{k+1} = 	X_k + \zeta_x, 	&&T_{k+1} = T_k + \tau_k,
\end{align}
where the initial position $X_0 = 0$, the initial time is $T_0=0$, $\zeta_x$ is the longitudinal (i.e., along the mean flow direction) correlation length, and $\tau_k$ is the transition time, or duration, associated with step $k$.

\begin{figure*}[]
	\includegraphics[width=0.45\textwidth]{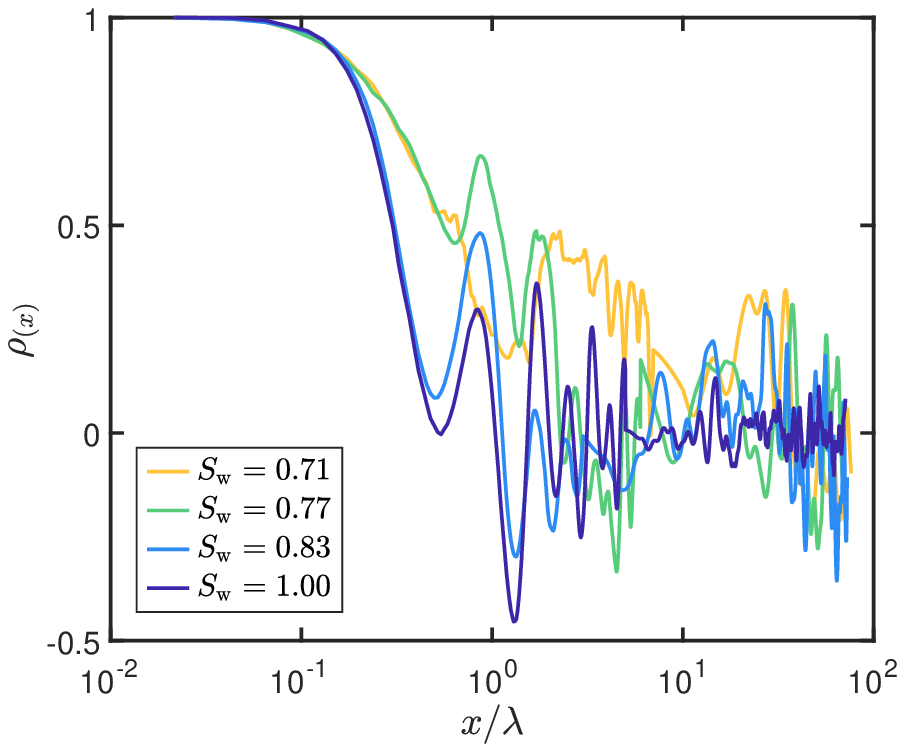}
	\caption{\label{fig:autocorrelation} Auto-correlation functions $\rho$ of the longitudinal velocities, as a function of longitudinal distance $x$ normalized by the mean pore size $\lambda$ for $S_{\mathrm{w}}=1.00$, $0.83$, $0.77$, and $0.71$.
	}
\end{figure*}

We determine $\zeta_x$ based on $s$-Lagrangian (i.e., sampled equidistantly along particle trajectories) velocity statistics obtained from linear interpolation of the $t$-Lagrangian trajectories. Correlation lengths increase with decreasing saturation; the values obtained are given in Table~\ref{tab:values_parameters}. The corresponding auto-correlation functions, from where the values of $\zeta_x$ for each saturation are obtained, are shown in Fig.~\ref{fig:autocorrelation}.

The velocities $V_k$ of a particle are assumed to be constant throughout each step $k$, and fully uncorrelated between steps, as an approximation of the Lagrangian correlation structure of the flow associated with the longitudinal correlation length $\zeta_x$. Accordingly, for each step $k$, $V_k$ is independently sampled from the equilibrium $s$-Lagrangian velocity PDF, which is given by $p_\mathrm{s}(v)= vp_E(v)/\bar v$. The flux-weighting of the underlying Eulerian PDF is required by the fixed-distance sampling of particle velocities~\cite{Dentz2016}. Note that the initial velocity $V_0$ is determined by the initial condition, which here is also flux-weighted. The transition time $\tau_k$ associated with each fixed longitudinal displacement depends on the current particle velocity and on its orientation with respect to the mean flow direction. The latter can be quantified in terms of the average tortuosity $\chi$, determined as the ratio of average Eulerian velocity magnitudes to the average projection of Eulerian velocities along the mean flow direction. Values of tortuosity for each analyzed saturation degree are reported in Table~\ref{tab:values_parameters}. The correlation length along streamlines, corresponding to the full particle displacement within a step, is then $\zeta=\chi\zeta_x$, and
\begin{equation}
\tau_k = \frac{\chi\zeta_x}{V_k}.
\end{equation}

\begin{figure*}[]
	\includegraphics{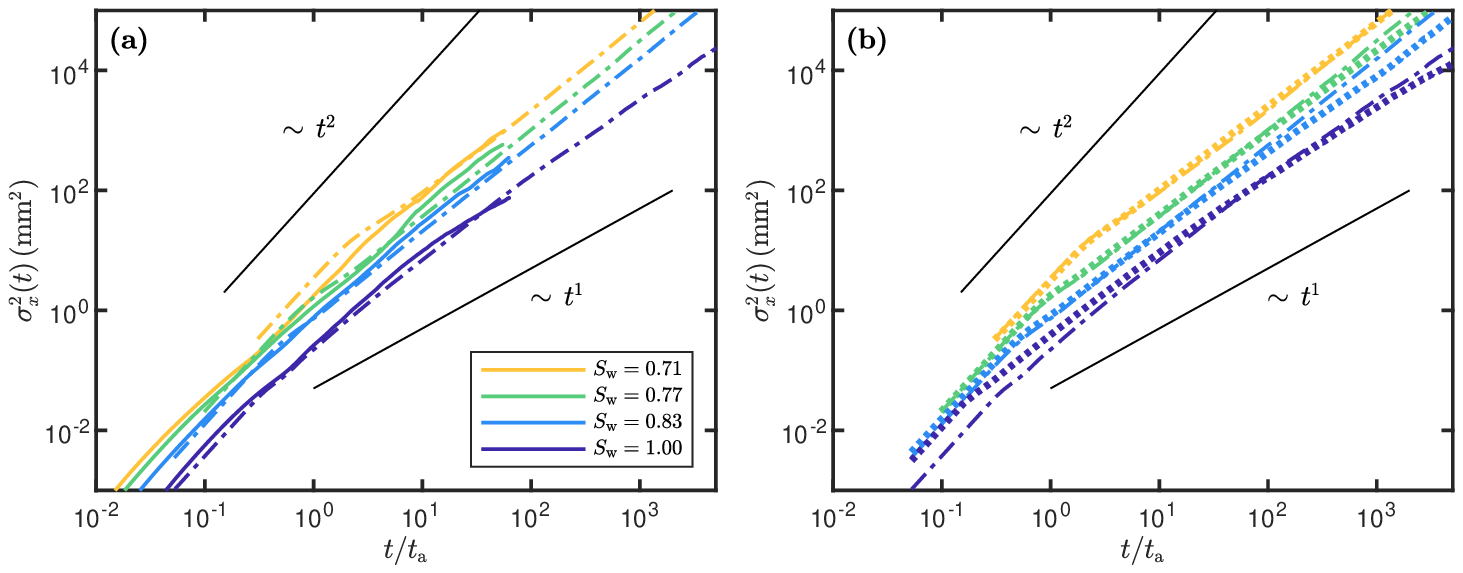}
	\caption{\label{fig:CTRW} Temporal evolution of advective longitudinal dispersion $\sigma^2_{x}$. Time has been non-dimensionalized by the advective time $t_{\mathrm{a}}$ for each saturation. (a) Comparison of dispersion from particle tracking simulations (continuous lines) with dispersion obtained from CTRW using the numerical Eulerian velocity PDFs (dash-dotted lines). (b) Comparison of $\sigma^2_{x}$ from CTRW using the numerical Eulerian velocity PDFs (dash-dotted lines) and from CTRW using the velocity PDFs from the theoretical model (dotted lines). Scalings for ballistic ($\sigma^2_{x} \sim t^2$) and Fickian ($\sigma^2_{x} \sim t^1$) regimes are shown as visual guides.
	}
\end{figure*}

As discussed in the main text, we employ the CTRW approach to compute the temporal evolution of the advective longitudinal dispersion $\sigma^2_{x}$ based on our predictions for the velocity PDF. Figure~\ref{fig:CTRW} shows additional comparisons to the results obtained from parameterizing the CTRW using the velocity PDF obtained directly from the simulated velocity field.

\bibliography{BibliographyPRL2020_20210314}